# Reduced Harmonic Representation of Partitions

*Michalis Psimopoulos*

*Plasma Group, Physics Department, Imperial College, London SW7 2BZ, UK*
*Email:m.psimopoulos@imperial.ac.uk*

**Abstract.** In the present paper the number of positive integer solutions of the equation $n_1 + 2n_2 + 3n_3 + \cdots + sn_s = s$ representing also the number of energy states of a conservative system of particles as well as the number of *partitions* of the positive integer *s*, is expressed in terms of harmonic functions by the *reduced* integral

$$p_s = \frac{2}{\pi} \int_0^{\pi/2} \prod_{\kappa=1}^{s} \left\{ \frac{\sin[(s+\kappa)x]}{\sin(\kappa x)} \right\} \cos[(s^2 - 2s)x] dx$$

It is shown that this result is due to the structure of the tail of the finite Fourier series of the kernel function $\mathcal{D}_s(x) = \prod_{\kappa=1}^{s} \{\sin[(s+\kappa)x]/\sin(\kappa x)\}$ which has the form
$2\{p_s \cos[(s^2 - 2s)x] + p_{s-1}\cos[(s^2 - 2(s-1))x] + \cdots + p_1 \cos[(s^2 - 2)x] + p_o \cos(s^2 x)\}$
where $p_o = 1$ and $p_1, p_2, \ldots, p_s$ are the partitions the of numbers $1, 2, \ldots, s$ respectively. Using the method of induction it is proved that $p_s$ can be also expressed by the general formula

$$p_s = \frac{2}{\pi} \int_0^{\pi/2} \mathcal{D}_{s+m}(x) \cos\{[(s+m)^2 - 2s]x\} dx \; ; \quad m = 0,1,2,3,\ldots$$

where the reduced integral is obtained for *m* = 0.

## 1. Introduction

In a previous article[1] it was argued that in the case of redistribution of $N$ particles between *s*+1 energy levels *0, ε, 2ε, 3ε,...,sε* where $N > s$ and *E=sε* is the total energy of the system, conservation of energy implies that the number of states of this system is equal to the number of integer solutions of the equation

$$n_1 + 2n_2 + 3n_3 + \cdots + sn_s = s \tag{1}$$

where $n_1 \geq 0, n_2 \geq 0, \ldots, n_s \geq 0$. On the other hand, if $n_1, n_2, \ldots, n_s$ represents respectively the number of times that the numbers 1,2,...,*s* occur in a certain partition of a positive integer *s*, then the number of partitions $p_s$ of *s* is also equal[2] to the number of integer solutions of the Eq.(1). From the geometrical point of view Eq.(1) represents an hyperplane in a *s*-dimensional cartesian space $[n_1, n_2, \ldots, n_s]$ cutting each axis at $n_1 = s; n_2 = s/2; n_3 = s/3; \ldots; n_s = 1$ respectively. In article[1], $p_s$ was expressed as a sum of characteristic functions over all integer mesh points of the hypercube: $0 \leq n_1 \leq s ; 0 \leq n_2 \leq s ; \ldots; 0 \leq n_s \leq s$ viz.

$$p_s = \sum_{n_1=0}^{s} \sum_{n_2=0}^{s} \cdots \sum_{n_s=0}^{s} \delta(n_1 + 2n_2 + 3n_3 + \cdots + sn_s - s) \tag{2}$$

leading to the harmonic representation

$$p_s = \frac{2}{\pi}\int_0^{\pi/2} \prod_{\kappa=1}^{s} \left\{\frac{\sin[\kappa(s+1)x]}{\sin(\kappa x)}\right\} \cos\left\{\left[\frac{s^2(s+1)}{2} - 2s\right]x\right\} dx \tag{3}$$

In the present work we express $p_s$ as

$$p_s = \sum_{n_1=0}^{s} \sum_{n_2=0}^{s/2} \sum_{n_3=0}^{s/3} \cdots \sum_{n_s=0}^{1} \delta(n_1 + 2n_2 + 3n_3 + \cdots + sn_s - s) \tag{4}$$

where

$$\delta(m-n) = \begin{cases} 1 & m = n \\ 0 & m \neq n \end{cases} \tag{5}$$

In this way we cover the hyperplane (1) with the *minimal* number of positive integer mesh points of the hyperspace $[n_1, n_2, \ldots, n_s]$. However, since the numbers $s/2, s/3, \ldots, s/(s-1)$ are not necessarily integers, we introduce the concept of a *fractional sum* of harmonic functions as an extension of an ordinary sum in the following sense: An ordinary sum of *sines* reads

$$\sum_{n=0}^{s} \sin(2n\kappa x + y) = \frac{\sin[(s+1)\kappa x]}{\sin(\kappa x)} \sin(s\kappa x + y) ; \quad \kappa = 1, 2, \ldots, s \tag{6a}$$

whereas a fractional sum of *sines* is defined as

$$\sum_{n=0}^{s/\kappa} \sin(2n\kappa x + y) = \frac{\sin[(s+\kappa)x]}{\sin(\kappa x)} \sin(sx + y) ; \quad \kappa = 1, 2, \ldots, s \tag{6b}$$

Similarly, an ordinary sum of *cosines* reads

$$\sum_{n=0}^{s} \cos(2n\kappa x + y) = \frac{\sin[(s+1)\kappa x]}{\sin(\kappa x)} \cos(s\kappa x + y) ; \quad \kappa = 1, 2, \ldots, s \tag{7a}$$

whereas a fractional sum of *cosines* is defined as

$$\sum_{n=0}^{s/\kappa} \cos(2n\kappa x + y) = \frac{\sin[(s+\kappa)x]}{\sin(\kappa x)} \cos(sx + y) ; \quad \kappa = 1, 2, \ldots, s \tag{7b}$$

In the present article we use the same method as in Ref.[1] in order to express $p_s$ as an integral over harmonic functions. However, since the concept of fractional sum is rather formal, additional analysis is needed in order to establish the validity of the results.

## 2. Reduced harmonic representation of partitions

There are two ways of expressing $\delta(m-n)$ in terms of orthogonal harmonic functions

I. $\quad \delta(m-n) = \frac{2}{\pi}\int_0^\pi \sin(mx)\sin(nx)dx$ (8)

so that by changing variable $x \to 2x$, Eq.(4) reads

$$p_s = \frac{4}{\pi}\int_0^{\pi/2} \sin(2sx)\left\{\sum_{n_1=0}^{s}\sum_{n_2=0}^{s/2}\sum_{n_3=0}^{s/3}\cdots\sum_{n_s=0}^{1} \sin[(n_1+2n_2+3n_3\ldots+sn_s)2x]\right\}dx \quad (9)$$

sum over $n_1$

$$\sum_{n_1=0}^{s}\sin[(n_1+2n_2+3n_3+\cdots+sn_s)2x]$$

$$= \frac{\sin[(s+1)x]}{\sin x}\sin[(2n_2+3n_3+\cdots+sn_s)2x+sx] \quad (10)$$

sum over $n_1, n_2$

$$\sum_{n_1=0}^{s}\sum_{n_2=0}^{s/2}\sin[(n_1+2n_2+3n_3+\cdots+sn_s)2x]$$

$$= \frac{\sin[(s+1)x]}{\sin x}\sum_{n_2=0}^{s/2}\sin[(2n_2+3n_3+\cdots+sn_s)2x+sx]$$

$$= \frac{\sin[(s+1)x]\sin[(s+2)x]}{\sin x \sin(2x)}\sin[(3n_3+\cdots+sn_s)2x+2sx] \quad (11)$$

sum over $n_1, n_2, n_3$

$$\sum_{n_1=0}^{s}\sum_{n_2=0}^{s/2}\sum_{n_3=0}^{s/3}\sin[(n_1+2n_2+3n_3\ldots+sn_s)2x]$$

$$= \frac{\sin[(s+1)x]\sin[(s+2)x]}{\sin x \sin(2x)}\sum_{n_3=0}^{s/3}\sin[(3n_3+4n_4+\cdots+sn_s)2x+2sx]$$

$$= \frac{\sin[(s+1)x]\sin[(s+2)x]\sin[(s+3)x]}{\sin x \sin(2x)\sin(3x)}\sin[(4n_4+\cdots+sn_s)2x+3sx] \quad (12)$$

.................................................................................................................................

sum over $n_1, n_2, n_3,\ldots, n_s$

$$\sum_{n_1=0}^{s}\sum_{n_2=0}^{s/2}\sum_{n_3=0}^{s/3}\ldots\sum_{n_s=0}^{1}\sin[(n_1+2n_2+3n_3\ldots+sn_s)2x]$$

$$=\frac{\sin[(s+1)x]\sin[(s+2)x]\sin[(s+3)x]\ldots\sin[(s+s)x]}{\sin x\sin(2x)\sin(3x)\ldots\sin(sx)}\sin(s^2x) \qquad (13)$$

Substituting the above result into Eq.(9) we get

$$p_s=\frac{4}{\pi}\int_0^{\pi/2}\frac{\sin[(s+1)x]\sin[(s+2)x]\sin[(s+3)x]\ldots\sin(2sx)}{\sin x\sin(2x)\sin(3x)\ldots\sin(sx)}\sin(2sx)\sin(s^2x)dx \qquad (14)$$

II. $\quad \delta(m-n)=\frac{2}{\pi}\int_0^{\pi}\cos(mx)\cos(nx)dx \qquad (15)$

so that by changing variable $x\to 2x$, Eq.(4) reads

$$p_s=\frac{4}{\pi}\int_0^{\pi/2}\cos(2sx)\left\{\sum_{n_1=0}^{s}\sum_{n_2=0}^{s/2}\sum_{n_3=0}^{s/3}\ldots\sum_{n_s=0}^{1}\cos[(n_1+2n_2+3n_3+\cdots+sn_s)2x]\right\}dx \qquad (16)$$

sum over $n_1$

$$\sum_{n_1=0}^{s}\cos[(n_1+2n_2+3n_3+\cdots+sn_s)2x]$$

$$=\frac{\sin[(s+1)x]}{\sin x}\cos[(2n_2+3n_3+\cdots+sn_s)2x+sx] \qquad (17)$$

sum over $n_1, n_2$

$$\sum_{n_1=0}^{s}\sum_{n_2=0}^{s/2}\cos[(n_1+2n_2+3n_3+\cdots+sn_s)2x]$$

$$=\frac{\sin[(s+1)x]}{\sin x}\sum_{n_2=0}^{s/2}\cos[(2n_2+3n_3+\cdots+sn_s)2x+sx]$$

$$=\frac{\sin[(s+1)x]\sin[(s+2)x]}{\sin x\sin(2x)}\cos[(3n_3+\cdots+sn_s)2x+2sx] \qquad (18)$$

sum over $n_1, n_2, n_3$

$$\sum_{n_1=0}^{s}\sum_{n_2=0}^{s/2}\sum_{n_3=0}^{s/3}\cos[(n_1+2n_2+3n_3+\cdots+sn_s)2x]$$

$$= \frac{\sin[(s+1)x]\sin[(s+2)x]}{\sin x \sin(2x)} \sum_{n_3=0}^{s/3} \cos[(3n_3 + 4n_4 + \cdots + sn_s)2x + 2sx]$$

$$= \frac{\sin[(s+1)x]\sin[(s+2)x]\sin[(s+3)x]}{\sin x \sin(2x)\sin(3x)} \cos[(4n_4 + \cdots + sn_s)2x + 3sx] \qquad (19)$$

................................................................................................................................

sum over $n_1, n_2, n_3, \ldots, n_s$

$$\sum_{n_1=0}^{s} \sum_{n_2=0}^{s/2} \sum_{n_3=0}^{s/3} \cdots \sum_{n_s=0}^{1} \cos[(n_1 + 2n_2 + 3n_3 + \cdots + sn_s)2x]$$

$$= \frac{\sin[(s+1)x]\sin[(s+2)x]\sin[(s+3)x]\ldots\sin[(s+s)x]}{\sin x \sin(2x) \sin(3x) \ldots \sin(sx)} \cos(s^2 x) \qquad (20)$$

Substituting the above result into Eq.(16) we get

$$p_s = \frac{4}{\pi} \int_0^{\pi/2} \frac{\sin[(s+1)x]\sin[(s+2)x]\sin[(s+3)x]\ldots\sin(2sx)}{\sin x \sin(2x)\sin(3x) \ldots \sin(sx)} \cos(2sx) \cos(s^2 x) dx \qquad (21)$$

Adding Eqs.(14,21) we obtain:

$$p_s = \frac{2}{\pi} \int_0^{\pi/2} \frac{\sin[(s+1)x]\sin[(s+2)x]\sin[(s+3)x]\ldots\sin(2sx)}{\sin x \sin(2x)\sin(3x) \ldots \sin(sx)} \cos[(s^2 - 2s)x] dx \qquad (22)$$

This is the *reduced* integral representation of $p_s$ in terms of harmonic functions. Clearly, Eq.(22) is a simpler formula than Eq.(3) obtained in Ref.[1]. But its derivation is rather formal as it is based on the concept of fractional sum. Therefore, further analysis will be presented in sections 3 and 4 in order to prove the above result.

### 3. Fourier series of the function $\mathcal{D}_s(x)$

In order to establish the validity of Eq.(22) we consider the function

$$\mathcal{D}_s(x) = \frac{\sin[(s+1)x]\sin[(s+2)x]\ldots\sin(2sx)}{\sin x \sin(2x) \ldots \sin(sx)} \qquad (23)$$

having the following properties

(i) $\quad \mathcal{D}_s(0) = \dfrac{(s+1)(s+2)\ldots(2s)}{1.2.\ldots.s} = \dfrac{(2s)!}{(s!)^2} = \binom{2s}{s} \qquad (24)$

(ii) $\quad \mathcal{D}_s(x) = \mathcal{D}_s(-x)$ is even function with Fourier series

$$\mathcal{D}_s(x) = \sum_{n=0}^{M} c_n \cos(nx) \tag{25}$$

where $M = s+1+s+2+\cdots+s+s-1-2-\cdots-s = s^2$ is the max. argument.

(iii) Consider the function

$$\mathcal{D}_s(x+\pi) = \frac{\sin[(s+1)(x+\pi)]\sin[(s+2)(x+\pi)]\ldots\sin[2s(x+\pi)]}{\sin(x+\pi)\sin[2(x+\pi)]\ldots\sin[s(x+\pi)]} \tag{26}$$

using $\sin[m(x+\pi)] = \sin(mx)\cos(m\pi) = (-1)^m \sin(mx)$, we have

$$\mathcal{D}_s(x+\pi) = \frac{(-1)^{s+1}(-1)^{s+2}\ldots(-1)^{2s}}{(-1)^1(-1)^2\ldots(-1)^s}\mathcal{D}_s(x) = (-1)^{s^2}\mathcal{D}_s(x) \tag{27}$$

and from $\cos[n(x+\pi)] = \cos(nx)\cos(n\pi) = (-1)^n \cos(nx)$, Eq.(25) reads

$$\mathcal{D}_s(x+\pi) = \sum_{n=0}^{M} c_n \cos[n(x+\pi)] = \sum_{n=0}^{M} (-1)^n c_n \cos(nx) \tag{28}$$

If $s=2\kappa$; $\kappa=1,2,3,\ldots$then $s^2$ is even and Eq.(27) implies $\mathcal{D}_s(x+\pi) = \mathcal{D}_s(x)$ so that comparing Eq.(25) with Eq.(28) we obtain $c_1=0, c_3=0,\ldots c_{M-1}=0$ and the Fourier series of $\mathcal{D}_s(x)$ has in this case the form:

$$\mathcal{D}_s(x) = a_o + a_1\cos 2x + a_2\cos 4x + \cdots + a_N \cos(s^2 x) \tag{29a}$$

where $N = s^2/2$; $a_o + a_1 + a_2 + \cdots + a_N = \binom{2s}{s}$ [see Eq.(24)].

If $s=2\kappa-1$; $\kappa=1,2,3,\ldots$then $s^2$ is odd and Eq.(27) implies $\mathcal{D}_s(x+\pi) = -\mathcal{D}_s(x)$ so that comparing Eq.(25) with Eq.(28) we obtain $c_0=0, c_2=0, c_4=0,\ldots,c_{M-1}=0$ and the Fourier series of $\mathcal{D}_s(x)$ has in this case the form:

$$\mathcal{D}_s(x) = a_1\cos x + a_2\cos 3x + a_3\cos 5x + \cdots + a_N \cos(s^2 x) \tag{29b}$$

where $N = (s^2+1)/2$; $a_1 + a_2 + a_3 + \cdots + a_N = \binom{2s}{s}$ [see Eq.(24)].

Using orthogonality of the harmonic functions, the coefficients of Eqs.(29) are given by

$$s = 2\kappa; \quad a_m = \frac{4}{\pi}\int_0^{\pi/2} \mathcal{D}_s(x)\cos(2mx)dx \; ; \quad m = 0,1,2,3,\ldots,\frac{s^2}{2} \tag{30a}$$

$$s = 2\kappa-1; \quad a_m = \frac{4}{\pi}\int_0^{\pi/2} \mathcal{D}_s(x)\cos[(2m-1)x]dx \; ; \quad m = 1,2,3,\ldots,\frac{s^2+1}{2} \tag{30b}$$

(iv) From definition (23) it is clear that $\mathcal{D}_s(x)$ and $\mathcal{D}_{s+1}(x)$ are related through

$$\mathcal{D}_{s+1}(x)\sin[(s+1)x] = \mathcal{D}_s(x)\{\sin[(3s+2)x] + \sin(sx)\} \tag{31}$$

using Eq.(31) as an *algorithm* we can obtain the Fourier series of the $\mathcal{D}_s(x)$ of various orders.

Example: for $s=1$; $\quad \mathcal{D}_1(x) = \frac{\sin 2x}{\sin x} = 2\cos x$ (32)

for $s=2$; $\quad \mathcal{D}_2(x) = \frac{\sin 3x \sin 4x}{\sin x \sin 2x} = 2(b_o + b_1\cos 2x + b_2\cos 4x)$

where $b_o, b_1, b_2$ are unknown coefficients.

Now Eq.(31) with $s=1$ reads

$2(b_o + b_1\cos 2x + b_2\cos 4x)\sin 2x = 2\cos x(\sin 5x + \sin x)$

$\Rightarrow (2b_o - b_2)\sin 2x + b_1\sin 4x + b_2\sin 6x = \sin 2x + \sin 4x + \sin 6x$

equating coefficients we get $b_o=1$; $b_1=1$; $b_2=1$ so that

$\mathcal{D}_2(x) = 2(1 + \cos 2x + \cos 4x)$ (33)

Clearly Eq.(22) is valid for both $\mathcal{D}_1(x)$; $\mathcal{D}_2(x)$ viz.

$p_1 = \frac{2}{\pi}\int_0^{\pi/2} \mathcal{D}_1(x)\cos x\, dx = \frac{4}{\pi}\int_0^{\pi/2} \cos^2 x\, dx = 1$;

$p_2 = \frac{2}{\pi}\int_0^{\pi/2} \mathcal{D}_2(x)\, dx = \frac{4}{\pi}\int_0^{\pi/2} (1 + \cos 2x + \cos 4x)\, dx = \frac{4}{\pi}\int_0^{\pi/2} dx = 2$ (34)

Using the same method we can derive the Fourier series of $\mathcal{D}_s(x)$ of any order. Below, we expand $\mathcal{D}_s(x)$ and validate Eq.(22) up to order $s=10$.

$s = 3$; $\quad \mathcal{D}_3(x) = \frac{\sin 4x \sin 5x \sin 6x}{\sin x \sin 2x \sin 3x}$ (35)

$\mathcal{D}_3(x) = 2\big(3\cos x + \underline{3\cos 3x + 2\cos 5x + \cos 7x + \cos 9x}\big)$ (36)

$t_3 = \frac{3^2+1}{2} = 5$ terms; $\mathcal{D}_3(0) = \binom{6}{3} = 20$; $\mathcal{D}_3\left(\frac{\pi}{2}\right) = 0$

Check: $\mathcal{D}_3(0) = 2(3 + 3 + 2 + 1 + 1) = 20$

$p_3 = \frac{2}{\pi}\int_0^{\pi/2} \mathcal{D}_3(x)\cos 3x\, dx = \frac{12}{\pi}\int_0^{\pi/2} \cos^2 3x\, dx = 3$ (37)

$s = 4$; $\quad \mathcal{D}_4(x) = \frac{\sin 5x \sin 6x \sin 7x \sin 8x}{\sin x \sin 2x \sin 3x \sin 4x}$ (38)

$\mathcal{D}_4(x) = 2(4 + 7\cos 2x + 7\cos 4x + 5\cos 6x$
$\qquad\qquad +5\cos 8x + 3\cos 10x + 2\cos 12x + \cos 14x + \cos 16x)$ (39)

$t_4 = \frac{4^2+2}{2} = 9$ terms; $\mathcal{D}_4(0) = \binom{8}{4} = 70$; $\mathcal{D}_4\left(\frac{\pi}{2}\right) = \binom{4}{2} = 6$

Check: $\mathcal{D}_4(0) = 2(4+7+7+5+5+3+2+1+1)=70$

$$p_4 = \frac{2}{\pi}\int_0^{\pi/2} \mathcal{D}_4(x)\cos 8x\, dx = \frac{20}{\pi}\int_0^{\pi/2}\cos^2 8x\, dx = 5 \tag{40}$$

$$s = 5; \quad \mathcal{D}_5(x) = \frac{\sin 6x \sin 7x \sin 8x \sin 9x \sin 10x}{\sin x \sin 2x \sin 3x \sin 4x \sin 5x} \tag{41}$$

$$\begin{aligned}\mathcal{D}_5(x) = 2(&20\cos x + 19\cos 3x + 18\cos 5x + 16\cos 7x + 14\cos 9x + 11\cos 11x + 9\cos 13x\\ &+7\cos 15x + 5\cos 17x + 3\cos 19x + 2\cos 21x + \cos 23x + \cos 25x)\end{aligned} \tag{42}$$

$$t_5 = \frac{5^2+1}{2} = 13 \text{ terms}; \quad \mathcal{D}_5(0) = \binom{10}{5} = 252; \quad \mathcal{D}_5\left(\frac{\pi}{2}\right) = 0$$

Check: $\mathcal{D}_5(0) = 2(20+19+18+16+14+11+9+7+5+3+2+1+1) = 252$

$$p_5 = \frac{2}{\pi}\int_0^{\pi/2} \mathcal{D}_5(x)\cos 15x\, dx = \frac{28}{\pi}\int_0^{\pi/2}\cos^2 15x\, dx = 7 \tag{43}$$

$$s = 6; \quad \mathcal{D}_6(x) = \frac{\sin 7x \sin 8x \sin 9x \sin 10x \sin 11x \sin 12x}{\sin x \sin 2x \sin 3x \sin 4x \sin 5x \sin 6x} \tag{44}$$

$$\begin{aligned}\mathcal{D}_6(x) = 2(&29 + 55\cos 2x + 55\cos 4x + 51\cos 6x + 48\cos 8x + 42\cos 10x + 39\cos 12x\\ &+32\cos 14x + 28\cos 16x + 22\cos 18x + 18\cos 20x + 13\cos 22x\\ &+11\cos 24x + 7\cos 26x + 5\cos 28x + 3\cos 30x + 2\cos 32x + \cos 34x + \cos 36x)\end{aligned} \tag{45}$$

$$t_6 = \frac{6^2+2}{2} = 19 \text{ terms}; \quad \mathcal{D}_6(x) = \binom{12}{6} = 924; \quad \mathcal{D}_6\left(\frac{\pi}{2}\right) = \binom{6}{3} = 20$$

Check: $\mathcal{D}_6(0) = 2(29+55+55+51+48+42+39+32+28+22+18\\+13+11+7+5+3+2+1+1) = 924$

$$p_6 = \frac{2}{\pi}\int_0^{\pi/2} \mathcal{D}_6(x)\cos 24x\, dx = \frac{44}{\pi}\int_0^{\pi/2}\cos^2 24x\, dx = 11 \tag{46}$$

$$s = 7; \quad \mathcal{D}_7(x) = \frac{\sin 8x \sin 9x \sin 10x \sin 11x \sin 12x \sin 13x \sin 14x}{\sin x \sin 2x \sin 3x \sin 4x \sin 5x \sin 6x \sin 7x} \tag{47}$$

$$\begin{aligned}\mathcal{D}_7(x) = 2(&169\cos x + 166\cos 3x + 162\cos 5x + 155\cos 7x + 146\cos 9x + 136\cos 11x\\ &+125\cos 13x + 112\cos 15x + 100\cos 17x + 87\cos 19x + 75\cos 21x + 63\cos 23x\\ &+53\cos 25x + 42\cos 27x + 34\cos 29x + 26\cos 31x + 20\cos 33x\\ &+15\cos 35x + 11\cos 37x + 7\cos 39x + 5\cos 41x + 3\cos 43x + 2\cos 45x\\ &+\cos 47x + \cos 49x)\end{aligned} \tag{48}$$

$$t_7 = \frac{7^2+1}{2} = 25 \text{ terms}; \quad \mathcal{D}_7(0) = \binom{14}{7} = 3432; \quad \mathcal{D}_7\left(\frac{\pi}{2}\right) = 0$$

Check: $\mathcal{D}_7(0) = 2(169+166+162+155+146+136+125+112+100+87+75\\+63+53+42+34+26+20+15+11+7+5+3+2+1+1) = 3432$

$$p_7 = \frac{2}{\pi}\int_0^{\pi/2} \mathcal{D}_7(x)\cos 35x\, dx = \frac{60}{\pi}\int_0^{\pi/2} \cos^2 35x\, dx = 15 \qquad (49)$$

$$s = 8;\quad \mathcal{D}_8(x) = \frac{\sin 9x \sin 10x \sin 11x \sin 12x \sin 13x \sin 14x \sin 15x \sin 16x}{\sin x \sin 2x \sin 3x \sin 4x \sin 5x \sin 6x \sin 7x \sin 8x} \qquad (50)$$

$$\begin{aligned}\mathcal{D}_8(x) = 2(&263 + 519\cos 2x + 515\cos 4x + 499\cos 6x + 486\cos 8x + 461\cos 10x \\ &+440\cos 12x + 409\cos 14x + 383\cos 16x + 348\cos 18x + 319\cos 20x \\ &+284\cos 22x + 255\cos 24x + 221\cos 26x + 194\cos 28x + 164\cos 30x \\ &+141\cos 32x + 116\cos 34x + 97\cos 36x + 77\cos 38x + 63\cos 40x + 48\cos 42x \\ &+38\cos 44x + 28\cos 46x + \underline{22\cos 48x + 15\cos 50x + 11\cos 52x} \\ &\underline{+7\cos 54x + 5\cos 56x + 3\cos 58x + 2\cos 60x + \cos 62x + \cos 64x})\end{aligned} \qquad (51)$$

$$t_8 = \frac{8^2+2}{2} = 33 \text{ terms};\quad \mathcal{D}_8(0) = \binom{16}{8} = 12870;\quad \mathcal{D}_8\left(\frac{\pi}{2}\right) = \binom{8}{4} = 70$$

Check: $\mathcal{D}_8(0) = 2(263+519+515+499+486+461+440+409+383+348+319+284$
$+255+221+194+164+141+116+97+77+63+48+38+28+22+15$
$+11+7+5+3+2+1+1)=12870$

$$p_8 = \frac{2}{\pi}\int_0^{\pi/2} \mathcal{D}_8(x)\cos 48x\, dx = \frac{88}{\pi}\int_0^{\pi/2} \cos^2 48x\, dx = 22 \qquad (52)$$

$$s = 9;\quad \mathcal{D}_9(x) = \frac{\sin 10x \sin 11x \sin 12x \sin 13x \sin 14x \sin 15x \sin 16x \sin 17x \sin 18x}{\sin x \sin 2x \sin 3x \sin 4x \sin 5x \sin 6x \sin 7x \sin 8x \sin 9x} \qquad (53)$$

$$\begin{aligned}\mathcal{D}_9(x) = 2(&1667\cos x + 1656\cos 3x + 1632\cos 5x + 1598\cos 7x + 1555\cos 9x \\ &+1499\cos 11x + 1437\cos 13x + 1368\cos 15x + 1292\cos 17x + 1210\cos 19x \\ &+1128\cos 21x + 1040\cos 23x + 954\cos 25x + 867\cos 27x + 782\cos 29x + 699\cos 31x \\ &+622\cos 33x + 545\cos 35x + 476\cos 37x + 411\cos 39x + 352\cos 41x + 297\cos 43x \\ &+251\cos 45x + 207\cos 47x + 171\cos 49x + 138\cos 51x + 111\cos 53x + 87\cos 55x \\ &+69\cos 57x + 52\cos 59x + 40\cos 61x + \underline{30\cos 63x + 22\cos 65x + 15\cos 67x} \\ &\underline{+11\cos 69x + 7\cos 71x + 5\cos 73x + 3\cos 75x + 2\cos 77x + \cos 79x + \cos 81x})\end{aligned} \qquad (54)$$

$$t_9 = \frac{9^2+1}{2} = 41 \text{ terms};\quad \mathcal{D}_9(0) = \binom{18}{9} = 48620;\quad \mathcal{D}_9\left(\frac{\pi}{2}\right) = 0$$

Check: $\mathcal{D}_9(0) = 2(1667+1656+1632+1598+1555+1499+1437+1368+1292$
$+1210+1128+1040+954+867+782+699+622+545+476+411$
$+352+297+251+207+171+138+111+87+69+52+40+30+22$
$+15+11+7+5+3+2+1+1)=48620$

$$p_9 = \frac{2}{\pi}\int_0^{\pi/2} \mathcal{D}_9(x)\cos 63x\, dx = \frac{120}{\pi}\int_0^{\pi/2} \cos^2 63x\, dx = 30 \qquad (55)$$

$$s = 10;\quad \mathcal{D}_{10}(x) = \frac{\sin 11x \sin 12x \sin 13x \sin 14x \sin 15x \sin 16x \sin 17x \sin 18x \sin 19x \sin 20x}{\sin x \sin 2x \sin 3x \sin 4x \sin 5x \sin 6x \sin 7x \sin 8x \sin 9x \sin 10x} \qquad (56)$$

$$\begin{aligned}\mathcal{D}_{10}(x) = 2(&2724 + 5424\cos 2x + 5392\cos 4x + 5311\cos 6x + 5226\cos 8x + 5095\cos 10x \\ &+4959\cos 12x + 4784\cos 14x + 4609\cos 16x + 4397\cos 18x + 4192\cos 20x\end{aligned}$$

$$+3956\cos22x + 3729\cos24x + 3481\cos26x + 3246\cos28x + 2994\cos30x$$
$$+2761\cos32x + 2517\cos34x + 2293\cos36x + 2065\cos38x + 1860\cos40x$$
$$+1652\cos42x + 1470\cos44x + 1289\cos46x + 1131\cos48x + 978\cos50x$$
$$+847\cos52x + 720\cos54x + 615\cos56x + 515\cos58x + 433\cos60x$$
$$+356\cos62x + 295\cos64x + 237\cos66x + +193\cos68x + 152\cos70x$$
$$+121\cos72x + 93\cos74x + 73\cos76x + 54\cos78x$$
$$\underline{+42\cos80x + 30\cos82x + 22\cos84x + 15\cos86x + 11\cos88x + 7\cos90x}$$
$$\underline{+5\cos92x + 3\cos94x + 2\cos96x + \cos98x + \cos100x)} \qquad (57)$$

$t_{10} = \frac{10^2+2}{2} = 51$ terms; $\mathcal{D}_{10}(0) = \binom{20}{10} = 184756$; $\mathcal{D}_{10}\left(\frac{\pi}{2}\right) = \binom{10}{5} = 252$

Check: $\mathcal{D}_{10}(0) = 2(2724+5424+5392+5311+5226+5095+4959+4784+4609+4397$
$+4192+3956+3729+3481+3246+2994+2761+2517+2293+2065$
$+1860+1652+1470+1289+1131+978+847+720+615+515+433$
$+356+295+237+193+152+121+93+73+54+42+30+22+15$
$+11+7+5+3+2+1+1)=184756$

$$p_{10} = \frac{2}{\pi}\int_0^{\pi/2} \mathcal{D}_{10}(x)\cos 80x\, dx = \frac{168}{\pi}\int_0^{\pi/2} \cos^2 80\, x\, dx = 42 \qquad (58)$$

We observe that the *tails* $\tau_s(x)$, of the above Fourier series of $\mathcal{D}_s(x)$, consisting of the $s+1$ leading terms with respect to argument and underlined in Eqs.(36,39,42,45,48,51,54,57), contain explicitly $p_o = 1$ and the partitions $p_1 = 1$, $p_2 = 2$, $p_3 = 3$, $p_4 = 5$, $p_5 = 7$, $p_6 = 11$, $p_7 = 15$, $p_8 = 22$, $p_9 = 30$, $p_{10} = 42$ as coefficients viz.

$$\tau_3(x) = 2(p_o\cos 9x + p_1\cos 7x + p_2\cos 5x + p_3\cos 3x)$$
$$\tau_4(x) = 2(p_o\cos 16x + p_1\cos 14x + p_2\cos 12x + p_3\cos 10x + p_4\cos 8x)$$
$$\tau_5(x) = 2(p_o\cos 25x + p_1\cos 23x + p_2\cos 21x + p_3\cos 19x + p_4\cos 17x + p_5\cos 15x)$$
$$\tau_6(x) = 2(p_o\cos 36x + p_1\cos 34x + p_2\cos 32x + p_3\cos 30x + p_4\cos 28x + p_5\cos 26x + p_6\cos 24x)$$
$$\tau_7(x) = 2(p_o\cos 49x + p_1\cos 47x + p_2\cos 45x + p_3\cos 43x + p_4\cos 41x + p_5\cos 39x$$
$$+p_6\cos 37x + p_7\cos 35x)$$
$$\tau_8(x) = 2(p_o\cos 64x + p_1\cos 62x + p_2\cos 60x + p_3\cos 58x + p_4\cos 56x + p_5\cos 54x$$
$$+p_6\cos 52x + p_7\cos 50x + p_8\cos 48x)$$
$$\tau_9(x) = 2(p_o\cos 81x + p_1\cos 79x + p_2\cos 77x + p_3\cos 75x + p_4\cos 73x + p_5\cos 71x$$
$$+p_6\cos 69x + p_7\cos 67x + p_8\cos 65x + p_9\cos 63x)$$
$$\tau_{10}(x) = 2(p_o\cos 100x + p_1\cos 98x + p_2\cos 96x + p_3\cos 94x + p_4\cos 92x + p_5\cos 90x$$
$$+p_6\cos 88x + p_7\cos 86x + p_8\cos 84x + p_9\cos 82x + p_{10}\cos 80x) \qquad (59)$$

From Eqs.(59) we expect that in general the tail of the Fourier series of $\mathcal{D}_s(x)$ has the form

$$\tau_s(x) = 2\{p_o\cos(s^2 x) + p_1\cos[(s^2-2)x] + p_2\cos[(s^2-4)x] + \cdots + p_s\cos[(s^2-2s)x]\} \qquad (60)$$

Clearly, Eq.(22) is a direct consequence of the structure of the last term $p_s\cos[(s^2-2s)x]$ of $\tau_s(x)$ is Eq.(60). Therefore, in order to validate Eq.(22) it is necessary to prove that the tail of $\mathcal{D}_s(x)$, has the form (60) for any $s=3,4,\ldots$.

## 4. Alternative derivation of the tail of $\mathcal{D}_s(x)$

We rewrite the function $\mathcal{D}_s(x)$; $s=3,4,\ldots,10$ given by Eqs.(35,38,41,44,47,50,53,56) as follows

$$\mathcal{D}_3(x) = \frac{-2^3 \sin x \sin 5x \cos 2x \cos 3x}{-2\sin^2 x}$$

$$\mathcal{D}_4(x) = \frac{-2^4 \sin x \sin 5x \sin 7x \cos 3x \cos 4x}{-2^2 \sin^2 x \sin 2x}$$

$$\mathcal{D}_5(x) = \frac{-2^5 \sin x \sin 7x \sin 9x \cos 3x \cos 4x \cos 5x}{-2^2 \sin^2 x \sin 2x}$$

$$\mathcal{D}_6(x) = \frac{2^6 \sin x \sin 7x \sin 9x \sin 11x \cos 4x \cos 5x \cos 6x}{2^3 \sin^2 x \sin 2x \sin 3x}$$

$$\mathcal{D}_7(x) = \frac{2^7 \sin x \sin 9x \sin 11x \sin 13x \cos 4x \cos 5x \cos 6x \cos 7x}{2^3 \sin^2 x \sin 2x \sin 3x}$$

$$\mathcal{D}_8(x) = \frac{2^8 \sin x \sin 9x \sin 11x \sin 13x \sin 15x \cos 5x \cos 6x \cos 7x \cos 8x}{2^4 \sin^2 x \sin 2x \sin 3x \sin 4x}$$

$$\mathcal{D}_9(x) = \frac{2^9 \sin x \sin 11x \sin 13x \sin 15x \sin 17x \cos 5x \cos 6x \cos 7x \cos 8x \cos 9x}{2^4 \sin^2 x \sin 2x \sin 3x \sin 4x}$$

$$\mathcal{D}_{10}(x) = \frac{-2^{10} \sin x \sin 11x \sin 13x \sin 15x \sin 17x \sin 19x \cos 6x \cos 7x \cos 8x \cos 9x \cos 10x}{-2^5 \sin^2 x \sin 2x \sin 3x \sin 4x \sin 5x} \quad (61)$$

Now, the tails of $\mathcal{D}_s(x)$ can be obtained if, we expand the numerators of $\mathcal{D}_s(x)$ in a Fourier series keeping leading terms with respect to argument, and then divide by the exact denominators. The numerators of $\mathcal{D}_s(x)$ given by Eq.(61) expand as follows

$a_3(x) = -2^3 \sin x \sin 5x \cos 2x \cos 3x \approx \cos 11x - \cos 9x + \cos 7x$

$a_4(x) = -2^4 \sin x \sin 5x \sin 7x \cos 3x \cos 4x \approx \sin 20x - \sin 18x + \sin 14x$

$a_5(x) = -2^5 \sin x \sin 7x \sin 9x \cos 3x \cos 4x \cos 5x \approx \sin 29x - \sin 27x + \sin 23x$

$a_6(x) = 2^6 \sin x \sin 7x \sin 9x \sin 11x \cos 4x \cos 5x \cos 6x \approx \cos 43x - \cos 41x + \cos 35x$

$a_7(x) = 2^7 \sin x \sin 9x \sin 11x \sin 13x \cos 4x \cos 5x \cos 6x \cos 7x \approx \cos 56x - \cos 54x + \cos 48x$

$a_8(x) = 2^8 \sin x \sin 9x \sin 11x \sin 13x \sin 15x \cos 5x \cos 6x \cos 7x \cos 8x \approx \sin 75x - \sin 73x + \sin 65x$

$a_9(x) = 2^9 \sin x \sin 11x \sin 13x \sin 15x \sin 17x \cos 5x \cos 6x \cos 7x \cos 8x \cos 9x$
$\quad \approx \sin 92x - \sin 90x + \sin 82x$

$a_{10}(x) = -2^{10} \sin x \sin 11x \sin 13x \sin 15x \sin 17x \sin 19x \cos 6x \cos 7x \cos 8x \cos 9x \cos 10x$
$\quad \approx \cos 116x - \cos 114x + \cos 104x \quad (62)$

The general formula of the above approximation of the numarators for arbitrary $s$ must be separated in four parts

(i) $\quad s = 4κ-1; κ = 1,2,3,\ldots \quad (s=3, 7, 11, 15,\ldots)$

$$a_s(x) = (-1)^{\frac{s+1}{4}} 2^s \sin x \sin[(s+2)x] \sin[(s+4)x] \ldots \sin[(2s-1)x]$$
$$\times \cos\left(\frac{s+1}{2}x\right)\cos\left(\frac{s+3}{2}x\right) \ldots \cos(sx)$$

$$a_s(x) \approx \cos\left(\frac{9s^2+7}{8}x\right) - \cos\left(\frac{9s^2-9}{8}x\right) + \cos\left(\frac{9s^2-8s-1}{8}x\right) \tag{63}$$

(ii) $\quad s = 4κ; κ = 1,2,3,\ldots \quad (s=4, 8, 12, 16,\ldots)$

$$a_s(x) = (-1)^{\frac{s}{4}} 2^s \sin x \sin[(s+1)x] \sin[(s+3)x] \ldots \sin[(2s-1)x]$$
$$\times \cos\left(\frac{s+2}{2}x\right)\cos\left(\frac{s+4}{2}x\right) \ldots \cos(sx)$$

$$a_s(x) \approx \sin\left(\frac{9s^2+2s+8}{8}x\right) - \sin\left(\frac{9s^2+2s-8}{8}x\right) + \sin\left(\frac{9s^2-6s-8}{8}x\right) \tag{64}$$

(iii) $\quad s = 4κ+1; κ = 1,2,3,\ldots \quad (s=5, 9, 13, 17,\ldots)$

$$a_s(x) = (-1)^{\frac{s-1}{4}} 2^s \sin x \sin[(s+2)x] \sin[(s+4)x] \ldots \sin[(2s-1)x]$$
$$\times \cos\left(\frac{s+1}{2}x\right)\cos\left(\frac{s+3}{2}x\right) \ldots \cos(sx) \tag{65}$$

$$a_s(x) \approx \sin\left(\frac{9s^2+7}{8}x\right) - \sin\left(\frac{9s^2-9}{8}x\right) + \sin\left(\frac{9s^2-8s-1}{8}x\right)$$

(iv) $\quad s = 4κ+2; κ = 1,2,3,\ldots \quad (s=6, 10, 14, 18,\ldots)$

$$a_s(x) = (-1)^{\frac{s+2}{4}} 2^s \sin x \sin[(s+1)x] \sin[(s+3)x] \ldots \sin[(2s-1)x]$$
$$\times \cos\left(\frac{s+2}{2}x\right)\cos\left(\frac{s+4}{2}x\right) \ldots \cos(sx) \tag{66}$$

$$a_s(x) \approx \cos\left(\frac{9s^2+2s+8}{8}x\right) - \cos\left(\frac{9s^2+2s-8}{8}x\right) + \cos\left(\frac{9s^2-6s-8}{8}x\right)$$

Using the above approximation for the numerators of $\mathcal{D}_s(x); s \geq 3$ in Eqs.(61) and keeping exact denominators we obtain

(i) $\quad s = 4κ-1; κ = 1,2,3,\ldots \quad (s=3, 7, 11, 15,\ldots)$

$$\mathcal{D}_s(x) \approx \frac{\cos\left(\frac{9s^2+7}{8}x\right) - \cos\left(\frac{9s^2-9}{8}x\right) + \cos\left(\frac{9s^2-8s-1}{8}x\right)}{(-1)^{\frac{s+1}{4}} 2^{\frac{s-1}{2}} \sin^2 x \sin 2x \sin 3x \sin 4x \ldots \sin\left(\frac{s-1}{2}x\right)} \tag{67}$$

(ii) $\quad s = 4κ; κ = 1,2,3,\ldots \quad (s=4, 8, 12, 16,\ldots)$

$$\mathcal{D}_s(x) \approx \frac{\sin\left(\frac{9s^2+2s+8}{8}x\right)-\sin\left(\frac{9s^2+2s-8}{8}x\right)+\sin\left(\frac{9s^2-6s-8}{8}x\right)}{(-1)^{\frac{s}{4}}2^{\frac{s}{2}}\sin^2 x\sin 2x\sin 3x\sin 4x\ldots\sin\left(\frac{s}{2}x\right)} \qquad (68)$$

(iii)  $s = 4\kappa + 1$ ; $\kappa = 1,2,3,\ldots$  ($s=5, 9, 13, 17,\ldots$)

$$\mathcal{D}_s(x) \approx \frac{\sin\left(\frac{9s^2+7}{8}x\right)-\sin\left(\frac{9s^2-9}{8}x\right)+\sin\left(\frac{9s^2-8s-1}{8}x\right)}{(-1)^{\frac{s-1}{4}}2^{\frac{s-1}{2}}\sin^2 x\sin 2x\sin 3x\sin 4x\ldots\sin\left(\frac{s-1}{2}x\right)} \qquad (69)$$

(iv)  $s = 4\kappa+2$; $\kappa = 1,2,3,\ldots$  ($s=6, 10, 14, 18,\ldots$)

$$\mathcal{D}_s(x) \approx \frac{\cos\left(\frac{9s^2+2s+8}{8}x\right)-\cos\left(\frac{9s^2+2s-8}{8}x\right)+\cos\left(\frac{9s^2-6s-8}{8}x\right)}{(-1)^{\frac{s+2}{4}}2^{\frac{s}{2}}\sin^2 x\sin 2x\sin 3x\sin 4x\ldots\sin\left(\frac{s}{2}x\right)} \qquad (70)$$

In particular, we can write the approximate expressions of $\mathcal{D}_s(x)$; $3 \leq s \leq 10$ as follows

$$\mathcal{D}_3(x) \approx \frac{\cos 11x - \cos 9x + \cos 7x}{-2\sin^2 x}$$

$$\mathcal{D}_4(x) \approx \frac{\sin 20x - \sin 18x + \sin 14x}{-2^2 \sin^2 x \sin 2x}$$

$$\mathcal{D}_5(x) \approx \frac{\sin 29x - \sin 27x + \sin 23x}{-2^2 \sin^2 x \sin 2x}$$

$$\mathcal{D}_6(x) \approx \frac{\cos 43x - \cos 41x + \cos 35x}{2^3 \sin^2 x \sin 2x \sin 3x}$$

$$\mathcal{D}_7(x) \approx \frac{\cos 56x - \cos 54x + \cos 48x}{2^3 \sin^2 x \sin 2x \sin 3x}$$

$$\mathcal{D}_8(x) \approx \frac{\sin 75x - \sin 73x + \sin 65x}{2^4 \sin^2 x \sin 2x \sin 3x \sin 4x}$$

$$\mathcal{D}_9(x) \approx \frac{\sin 92x - \sin 90x + \sin 82x}{2^4 \sin^2 x \sin 2x \sin 3x \sin 4x}$$

$$\mathcal{D}_{10}(x) \approx \frac{\cos 116x - \cos 114x + \cos 104x}{-2^5 \sin^2 x \sin 2x \sin 3x \sin 4x \sin 5x} \qquad (71)$$

Performing the above divisions it is easy to obtain in leading order with respect to arguments, the tails $\tau_s(x)$ of $\mathcal{D}_s(x)$ given by Eqs.(59) and containing $p_o$ and the partitions $p_1, p_2, \ldots, p_s$. It is interesting to observe however that only three leading terms of the numerator of $\mathcal{D}_s(x)$ existing in all expressions (71), are enough to provide by division the full tails $\tau_s(x)$. Inversely, by multiplying $\tau_s(x)$ with the exact denominators of $\mathcal{D}_s(x)$ in Eqs.(71), the partitions occurring within $\tau_s(x)$ should combine in such a way that, in leading order with respect to arguments, they should provide the simple coefficients +1,-1, +1 of the three leading terms of the numerators of $\mathcal{D}_s(x)$ in Eqs. (71).

In particular, using Eqs. (59) we see that this indeed happens:

$-2\sin^2 x \tau_3(x) = -4\sin^2 x(p_o\cos 9x + p_1\cos 7x + p_2\cos 5x + p_3\cos 3x)$

$\qquad = p_o\cos 11x + (p_1 - 2p_o)\cos 9x + (p_2 - 2p_1 + p_o)\cos 7x + \cdots$

Leading terms: $\cos 11x - \cos 9x + \cos 7x$ \hfill (72)

$-2^2\sin^2 x \sin 2x \tau_4(x)$

$\qquad = -8\sin^2 x \sin 2x(p_o\cos 16x + p_1\cos 14x + p_2\cos 12x + p_3\cos 10x + p_4\cos 8x)$

$\qquad = p_o\sin 20x + (p_1 - 2p_o)\sin 18x + (p_2 - 2p_1)\sin 16x + (p_3 - 2p_2 + 2p_o)\sin 14x + \cdots$

Leading terms: $\sin 20x - \sin 18x + \sin 14x$ \hfill (73)

$-2^2\sin^2 x \sin 2x \tau_5(x)$

$\qquad = -8\sin^2 x \sin 2x(p_o\cos 25x + p_1\cos 23x + p_2\cos 21x + p_3\cos 19x + p_4\cos 17x + p_5\cos 15x)$

$\qquad = p_o\sin 29x + (p_1 - 2p_o)\sin 27x + (p_2 - 2p_1)\sin 25x + (p_3 - 2p_2 + 2p_o)\sin 23x + \cdots$

Leading terms: $\sin 29x - \sin 27x + \sin 23x$ \hfill (74)

$2^3\sin^2 x \sin 2x \sin 3x \tau_6(x)$

$\qquad = 16\sin^2 x \sin 2x \sin 3x(p_o\cos 36x + p_1\cos 34x + p_2\cos 32x + p_3\cos 30x$
$\qquad\qquad + p_4\cos 28x + p_5\cos 26x + p_6\cos 24x)$

$\qquad = p_o\cos 43x + (p_1 - 2p_o)\cos 41x + (p_2 - 2p_1)\cos 39x$
$\qquad\qquad + (p_3 - 2p_2 + p_o)\cos 37x + (p_4 - 2p_3 + p_1 + p_o)\cos 35x + \cdots$

Leading terms: $\cos 43x - \cos 41x + \cos 35x$ \hfill (75)

$2^3\sin^2 x \sin 2x \sin 3x \tau_7(x)$

$\qquad = 16\sin^2 x \sin 2x \sin 3x(p_o\cos 49x + p_1\cos 47x + p_2\cos 45x + p_3\cos 43x + p_4\cos 41x +$
$\qquad\qquad p_5\cos 39x + p_6\cos 37x + p_7\cos 35x)$

$\qquad = p_o\cos 56x + (p_1 - 2p_o)\cos 54x + (p_2 - 2p_1)\cos 52x + (p_3 - 2p_2 + p_o)\cos 50x$
$\qquad\qquad + (p_4 - 2p_3 + p_1 + p_o)\cos 48x + \cdots$

Leading terms: $\cos 56x - \cos 54x + \cos 48x$ \hfill (76)

$2^4\sin^2 x \sin 2x \sin 3x \sin 4x \tau_8(x)$

$\qquad = 32\sin^2 x \sin 2x \sin 3x \sin 4x(p_o\cos 64x + p_1\cos 62x + p_2\cos 60x + p_3\cos 58x$
$\qquad\qquad + p_4\cos 56x + p_5\cos 54x + p_6\cos 52x + p_7\cos 50x + p_8\cos 48x)$

$\qquad = p_o\sin 75x + (p_1 - 2p_o)\sin 73x + (p_2 - 2p_1)\sin 71x + (p_3 - 2p_2 + p_o)\sin 69x$
$\qquad\qquad + (p_4 - 2p_3 + p_1)\sin 67x + (p_5 - 2p_4 + p_2 + 2p_o)\sin 65x + \cdots$

Leading terms: $\sin 75x - \sin 73x + \sin 65x$ \hfill (77)

$2^4\sin^2 x \sin 2x \sin 3x \sin 4x \tau_9(x)$

$\qquad = 32\sin^2 x \sin 2x \sin 3x \sin 4x(p_o\cos 81x + p_1\cos 79x + p_2\cos 77x + p_3\cos 75x$
$\qquad\qquad + p_4\cos 73x + p_5\cos 71x + p_6\cos 69x + p_7\cos 67x + p_8\cos 65x + p_9\cos 63x)$

$$= p_o\sin 92x + (p_1 - 2p_o)\sin 90x + (p_2 - 2p_1)\sin 88x + (p_3 - 2p_2 + p_o)\sin 86x$$
$$+ (p_4 - 2p_3 + p_1)\sin 84x + (p_5 - 2p_4 + p_2 + 2p_o)\sin 82x + \cdots$$

Leading terms: $\sin 92x - \sin 90x + \sin 82x$                                                              (78)

$-2^5 \sin^2 x \sin 2x \sin 3x \sin 4x \sin 5x \tau_{10}(x)$

$$= 64\sin^2 x \sin 2x \sin 3x \sin 4x \sin 5x (p_o\cos 100x + p_1\cos 98x + p_2\cos 96x + p_3\cos 94x +$$
$$p_4\cos 92x + p_5\cos 90x + p_6\cos 88x + p_7\cos 86x + p_8\cos 84x + p_9\cos 82x + p_{10}\cos 80x)$$

$$= p_o\cos 116x + (p_1 - 2p_o)\cos 114x + (p_2 - 2p_1)\cos 112x + (p_3 - 2p_2 + p_o)\cos 110x$$
$$+ (p_4 - 2p_3 + p_1)\cos 108x + (p_5 - 2p_4 + p_2 + p_o)\cos 106x$$
$$+ (p_6 - 2p_5 + p_3 + p_1)\cos 104x \ldots$$

Leading terms: $\cos 116x - \cos 114x + \cos 104x$                                           (79)

We observe that the numerators of $\mathcal{D}_s(x)$ in Eqs. (71) are indeed reproduced exactly by the above products. In particular, the coefficients of the first two leading terms are rather simple: $p_o = +1$; $p_1 - 2p_o = -1$ and remain the same in all products (arbitrary $s$). However, the coefficient of the third leading term is equal to +1 but its structure is rather involved and is given for various $s$ by the table

$s = 3$        $p_2 - 2p_1 + p_o = 1$

$s = 4,5$     $p_3 - 2p_2 + 2p_o = 1$

$s = 6,7$     $p_4 - 2p_3 + p_1 + p_o = 1$                                                         (80)

$s = 8,9$     $p_5 - 2p_4 + p_2 + 2p_o = 1$

$s = 10,11$   $p_6 - 2p_5 + p_3 + p_1 = 1$

The algorithm behind table (80) becomes clear if each coefficient is expressed as a difference between two successive partitions in terms of the *Euler recursion formula*[2]:

$s = 3$        $p_2 - 2p_1 + p_o = (p_2 - p_1 - p_o) - (p_1 - p_o) + p_o$

$s = 4,5$     $p_3 - 2p_2 + 2p_o = (p_3 - p_2 - p_1) - (p_2 - p_1 - p_o) + p_o$

$s = 6,7$     $p_4 - 2p_3 + p_1 + p_o = (p_4 - p_3 - p_2) - (p_3 - p_2 - p_1) + p_o$               (81)

$s = 8,9$     $p_5 - 2p_4 + p_2 + 2p_o = (p_5 - p_4 - p_3 + p_o) - (p_4 - p_3 - p_2) + p_o$

$s = 10,11$   $p_6 - 2p_5 + p_3 + p_1 = (p_6 - p_5 - p_4 + p_1) - (p_5 - p_4 - p_3 + p_o) + p_o$

Hence, for arbitrary $s$ the coefficients of the third leading term for two successive products

$s = 2\kappa - 2$; $s = 2\kappa - 1$ reads

$$[p_\kappa - (p_{\kappa-1} + p_{\kappa-2} - p_{\kappa-5} - p_{\kappa-7} + \cdots)] - [p_{\kappa-1} - (p_{\kappa-2} + p_{\kappa-3} - p_{\kappa-6} - p_{\kappa-8} + \cdots)] + p_o$$

$$= (p_\kappa - 2p_{\kappa-1} + p_{\kappa-3} + p_{\kappa-5} - p_{\kappa-6} + p_{\kappa-7} - p_{\kappa-8} - p_{\kappa-12} + p_{\kappa-13} - p_{\kappa-15} + \cdots) + p_o = 1 \quad (82)$$

Eq. (82) reproduces all the terms of table (80) and beyond for κ=2,3,4,... . Hence, the tail of $\mathcal{D}_s(x)$ has the form (60) and in turn the main result expressed by Eq. (22) is valid.

## 5. Generalization of the reduced integral

In section 3 it was proved that the function $\mathcal{D}_s(x)$, defined by Eq.(23), can be expanded in two finite cosine Fourier series given by Eqs.(29a,29b) where in both cases the term of maximum argument is $\cos(s^2 x)$. Hence for every $s=1,2,3...$ we have

$$\int_0^{\pi/2} \mathcal{D}_s(x)\cos[(s^2 + m)x]dx = 0 \ ; \quad m = 1,2,3,... \tag{83}$$

Using Eq.(83) we can derive the following formula

$$p_s = \frac{2}{\pi}\int_0^{\pi/2} \mathcal{D}_{s+m}(x)\cos\{[(s + m)^2 - 2s]x\}dx \ ; \quad m = 0,1,2,3,... \tag{84}$$

The proof of Eq.(84) will be based on the method of *induction*.

I. For $m=0$, Eq.(84) coincides with representation (22) and therefore it is valid.

II. Assume Eq.(84) is true for $m=κ$ viz.

$$p_s = \frac{2}{\pi}\int_0^{\pi/2} \mathcal{D}_{s+κ}(x)\cos\{[(s + κ)^2 - 2s]x\}dx \tag{85}$$

and prove that it is also true for $m=κ+1$ viz.

$$p_s = \frac{2}{\pi}\int_0^{\pi/2} \mathcal{D}_{s+κ+1}(x)\cos\{[(s + κ + 1)^2 - 2s]x\}dx \tag{86}$$

Proof: Consider Eq.(31) where $s$ is replaced by $s+κ$

$$\mathcal{D}_{s+κ+1}(x)\sin[(s + κ + 1)x] = \mathcal{D}_{s+κ}(x)\{\sin[(3s + 3κ + 2)x] + \sin[(s + κ)x]\} \tag{87}$$

multiply both sides of Eq.(87) by $2\sin\{[(s+κ)^2+s+3κ+2]x\}$

LHS $= 2\sin\{[(s + κ)^2 + s + 3κ + 2]x\}\sin[(s + κ + 1)x]\mathcal{D}_{s+κ+1}(x)$

$\quad = \{\cos\{[(s + κ + 1)^2 - 2s]x\} - \cos\{[(s + κ + 1)^2 + 2κ + 2]x\}\}\mathcal{D}_{s+κ+1}(x)$

RHS $= 2\sin\{[(s + κ)^2 + s + 3κ + 2]x\}\{\sin[(3s + 3κ + 2)x] + \sin[(s + κ)x]\}\mathcal{D}_{s+κ}(x)$

$\quad = \{\cos\{[(s + κ)^2 - 2s]x\} - \cos\{[(s + κ)^2 + 4s + 6κ + 4]x\}$

$\qquad +\cos\{[(s + κ)^2 + 2κ + 2]x\} - \cos\{[(s + κ)^2 + 2s + 4κ + 2]x\}\}\mathcal{D}_{s+κ}(x) \tag{88}$

integrating both sides and using Eqs.(83) we obtain

$$\frac{2}{\pi}\int_0^{\pi/2} \mathcal{D}_{s+\kappa+1}(x)\cos\{[(s+\kappa+1)^2 - 2s]x\}dx = \frac{2}{\pi}\int_0^{\pi/2} \mathcal{D}_{s+\kappa}(x)\cos\{[(s+\kappa)^2 - 2s]x\}dx \qquad (89)$$

according to Eq.(85) the RHS of Eq.(89) is equal to $p_s$, therefore the LHS of Eq.(89) is also equal to $p_s$ and Eq.(86) is true. This completes the proof of Eq.(84) by induction. Formula (84) is a generalization of the reduced integral representation (22) of $p_s$ in terms of harmonic functions.

## 6. Conclusion

In the present article the reduced integral representation of partitions in terms of harmonic products has been derived first by using hypergeometry and the new concept of fractional sum defined by Eqs.(6b,7b) and secondly by studying the Fourier series of the function $\mathcal{D}_s(x)$ appearing as a kernel in the integrals and defined by Eq.(23). The basic result obtained is given by Eq.(22) and its generalization by Eq.(84). We observe that formula (22) is simpler than the harmonic representation of $p_s$ given by Eq.(3), obtained in Ref.[1]. However, its derivation is certainly more involved and led to new relations between partitions and harmonic functions.

## References

[1] M. Psimopoulos, Harmonic Representation of Combinations and Partitions, arXiv:1102.5674v2 [math-ph].

[2] T. M. Apostol, Introduction to Analytic Number Theory, Spinger-Verlag, New York (1976).